\input{psfig}  
\def\Oiii{{[O{\,\small III}]~$\lambda$5007}\/}

\def\O4363{[O{\,\small III}]~$\lambda$4363\/}
\def\Nii{[N{\,\small II}]}
 \def\kms{{km~s$^{-1}$}\/}

\def\kms{{km~s$^{-1}$}\/}

\def\ergs.cm2.s{{ergs~cm$^{-2}$~s$^{-1}$}\/} 
 
\def\deg{{^{\circ}\!\!}\/} %\def\deg{{$^{\circ}$}\/}
\def\arcdeg{{^{\circ}\!\!}\/} %\def\deg{{$^{\circ}$}\/}

\newcommand{\HI}{H{\,\small I}}

\def\etal{{et~al.}}

\def\smallskip{\vskip 6pt}

\documentclass{aa}
\begin{document}
\thesaurus{ 06() }

\title{Radio Continuum Morphology of Southern \\ Seyfert Galaxies
\thanks{Based on observations done with the Australia Telescope
Compact Array (ATCA) and the Very Large Array(VLA)}}

\author{R.  Morganti\inst{1}\thanks{rmorgant@ira.bo.cnr.it}, Z.I. 
Tsvetanov\inst{2}, J.  Gallimore\inst{3}\thanks{Jansky Fellow}, M.G. 
Allen\inst{4}}

\institute{Istituto di Radioastronomia, CNR, Via Gobetti
101, 40129 Bologna, Italy \and 
Department of Physics and Astronomy, Johns Hopkins University,
Baltimore, MD 21218, USA \and
NRAO, 520 Edgemont Rd., Charlottesville, VA, 22903, USA
\and
Mount Stromlo and Siding Spring Observatories, 
Private Bag, Weston Creek Post Office, ACT 2611, Australia }

\offprints{R.Morganti}

\maketitle

\markboth{Radio Morphology of Southern Seyferts}{Morganti et al.}

\begin{abstract} We present radio observations for 29 southern Seyfert
galaxies selected from a volume limited sample with $cz<3600$ km s$^{-1}$, and
declination $\delta<0\degr$.  Objects with declination
$-30\degr<\delta<0\degr$ were observed with the Very Large Array (VLA) at 6~cm
(4.9 GHz) and objects with $\delta<-30\degr$ were observed with the Australia
Telescope Compact Array (ATCA) at 3.5~cm (8.6 GHz).  Both the VLA and the ATCA
observations have a resolution of $\sim 1\arcsec$.  These new observations
cover more than 50\% of the southern sample with all but two of the 29 objects
detected above our limit of 0.15 mJy.  Combining these data with data
available from the literature gives almost 85\% coverage of the southern
sample and further expands the radio observations of a distance limited sample
by 22\%. 

Collecting additional sources from the literature, consisting of known
Seyferts with $cz < 4600$~km~s$^{-1}$, we find that 38\% of the sources are
unresolved at arcsecond resolution, and 34\% have linear radio structure.  Our
results expand upon and are consistent with earlier studies.  We confirm a
correlation between the size of the radio structure and the radio luminosity. 
A comparison between Seyfert types 1 and 2 finds that type 2s tend to have a
larger linear size.  There is no statistically significant difference in radio
power between types 1 and 2, although all the most powerful objects appear to
be Seyfert 2's.  No significant difference has been found in the spectral
indices.

\end{abstract}


\section {Introduction} 

Radio emission from Seyfert galaxies traces energetic, mechanical processes
occurring in the active nucleus.  The radio continuum spectra of Seyferts are
almost invariably steep, consistent with optically-thin synchrotron emission. 
In those sources where the radio continuum has been resolved, the morphology
is commonly linear, interpreted to trace a stream of ejected plasma, or a jet,
originating from the central engine.  For example, most well-resolved Seyfert
radio sources are double or triple sources straddling the active nucleus.  The
radio emission of the nearest and brightest Seyferts resolves into tightly
collimated structures originating from the active nucleus, consistent with a
jet interpretation (e.g., NGC~1068: Gallimore et al.  1996; Muxlow et al. 
1996; Wilson \& Ulvestad  1983; NGC~4151: Pedlar et al. 1993; Mrk~3: Kukula et
al.  1993).  There is evidence that Seyfert radio jets impact the dense gas in
the near-nucleus environment and affect the gas distribution, morphology, and
ionization of the narrow-line region (NLR).  The orientation of the radio jet
is also the only straightforward measure of the symmetry axis of the active
nucleus, useful for testing orientation-based unifying schemes for AGNs. 

Early studies searching for differences between Seyfert 1's and 2's
(\cite{deb78} 1978; \cite{meu84}; \cite{ulv84a}, \cite{ulv84b})
concluded that Seyfert 2's have stronger and larger radio sources than
Seyfert 1's, however, these studies were biased by optical selection:
weaker Seyfert 2 galaxies were omitted from the samples.  There was
found to be little or no difference between Seyfert 1 and 2 radio
sources in follow-up studies which properly considered relatively
unbiased, volume-limited samples (\cite{ede87}; \cite{ulv89};
\cite{giu90} 1990).  More specifically, there is no statistically
significant difference in the distribution of radio luminosity, and
only a marginal difference in the distribution of radio source sizes,
with Seyfert 2's tending to be slightly larger than Seyfert 1's
(although only at $< 90$ \% significance, Wilson 1991).  In
contrast, unifying schemes predict that the radio jets in Seyfert 2's
should appear larger in projection, since the unifying model orients
the collimating disk more nearly edge-on in narrow-line AGNs.  It is
not simple to reconcile the statistics of radio sources with the
unified scheme hypothesis.

While there is no tendency for the radio sources to orient in any
preferred direction with respect to the plane of the host galaxy
(\cite{ulv84a}), the radio sources are commonly elongated in the same
direction as the NLR (\cite{han88} 1988; \cite{pog89};
\cite{wil88} 1988), i.e.  the inner part of the optical emission
regions (extending up to few kpc).  In some Seyfert~2s the NLR has a
conical or bi-conical shape with the nucleus at the apex (eg.
\cite{pog89}; Pogge 1997).  \cite{wil94} (1994) showed that the radio
axis is invariably co-aligned with the ionization cone axis for the 11
ionization cones known at that time. 

The alignment of the radio jet and NLR raises a question of energetics
--- does the radio jet ionize and heat the NLR significantly compared
to ionizing radiation from the AGN? In several well-studied cases,
there is a detailed morphological association between the NLR and the
radio emission (e.g.  Whittle et al. 1988, Whittle 1989, Capetti et
al. 1996; Gallimore et al. 1996). Currently, the numbers of such
well-studied cases is too small to address questions of interaction
and energetics in a statistically complete and meaningful sense. 

To this end, Tsvetanov et al.  (in prep.) have recently assembled a
volume-limited sample ($cz<3600$ \kms) of well-classified Seyfert galaxies. 
The main advantage of this sample is that all of the sources have been
extensively observed in the optical with narrow-band imaging done with the ESO
NTT, 3.6m and 2.2m telescopes.  This survey provides high-quality
emission-line maps (in \Oiii and H$\alpha$+\Nii) with a typical resolution of
$\sim1$\arcsec.  The sample includes objects with and without known extended
emission line regions, providing detailed information about the extent,
morphology, and degree of ionization (obtained from the \Oiii/H$\alpha$+\Nii
ratio) of the emission-line regions.  It is therefore important to obtain for
such a sample detailed radio images in order to perform a detailed comparison
of the radio and optical morphology on arcsecond (hundred-pc) scales. 

Here we present a radio imaging survey of the Tsvetanov et al.  sample, made
using the Very Large Array (VLA)\footnote{The National Radio Astronomy
Observatory is operated by Associated Universities, Inc., under contract with
National Science Foundation.} and the Australia Telescope Compact Array
(ATCA)\footnote{Operated by the CSIRO Australia Telescope National Facility.}
aperture synthesis telescopes.  Observing configurations and frequencies were
chosen to match the $\sim$ 1\arcsec\ resolution of the optical narrow-band
observations.  We discuss the radio properties and statistics for the 29
surveyed sources, and a more detailed comparison of the radio and optical
properties will be presented in a forthcoming paper.

\section {Observations}\label{observations}

\subsection{Sample}

The sample includes 51 well classified (from their optical spectra) Seyfert
galaxies south of declination $0\degr$ and with redshift $cz<3600$ \kms.  More
details about selection and optical observations of this sample are presented
in Tsvetanov et al. (in prep.), Tsvetanov et al.  (1998). 

The 29 sources listed in Table~1 were observed with either the VLA or
ATCA, depending on their declination. One source, NGC~3393, was  observed
with both arrays.  The array configurations and frequencies were
chosen to obtain a resolution of $\sim1$\arcsec\ in order to match the
resolution of the optical data.  In particular, the VLA was used in
its B and B$_N$A configurations, providing a resolution of $\sim$ 1
\arcsec\ at 6 cm (4.9~GHz).  To obtain similar resolution with the
ATCA required an observing frequency of 3.5~cm (8.6~GHz) with the 6 km
configuration.  Owing to missing short baselines in either array, our
data are more sensitive to compact emission from the circumnuclear
region but less sensitive to extended, diffuse emission. In addition,
faint, diffuse emission from active spirals tends to have very steeply
falling radio spectra with increasing frequency; at observing
wavelengths of 3.5 and 6~cm any diffuse emission will probably have
dropped below our surface brightness detection limit.

\begin{table*}
\voffset=1truecm
\begin{center}
{\bf Table 1:  VLA observations (6~cm)}

\medskip
\def\mc{\multicolumn{1}{c}}
\def\mcc{\multicolumn{2}{c}}
\def\mccc{\multicolumn{3}{c}}

\begin{tabular}{lrrrrrr@{}lr@{}lcc} \hline 
 Object   & \mc{${\alpha_{2000}}$} & \mc{${\delta_{2000}}$} &   
\mccc{Resolution} &  \mcc{rms} & \mcc{Peak} &  Array \\ 
          &  &      &  arcsec  & arcsec     & deg     &  \mcc{mJy beam$^{-1}$}     & 
\mcc{mJy beam$^{-1}$}  &   \\
\hline
          &         &           &       &        &           &         &           &   \\
NGC~1097  & 02 46 19.1 & $-$30 16 28 & 2.1   &  1.3   &  $-$59      & 0&.051   &    3&.78   &  B$_N$A \\
NGC~1320  & 03 24 48.7 & $-$03 02 33 & 2.5   &  1.4   &  $-$38      & 0&.081   &    1&.98   &  B\\
NGC~3393  & 10 48 24.0 & $-$25 09 40 & 2.2   &  1.3   &  $-$3       & 0&.049   &   15&.80   &  B$_N$A \\
NGC~3660  & 11 23 32.2 & $-$08 39 30 & 2.1   &  1.4   &  $-$1       & 0&.049   &    0&.49   &   B  \\
NGC~4968  & 13 07 06.0 & $-$23 40 43 & 2.3   &  1.4   &  $-$4       & 0&.046   &   13&.06   &  B$_N$A    \\
NGC~5427  & 14 03 25.9 & $-$06 01 50 & 2.1   &  1.4   &  $-$17      & 0&.048   &    2&.49   &  B    \\
NGC~7172  & 22 02 02.1 & $-$31 52 12 & 3.6   &  1.0   &   36      & 0&.053   &    3&.62   &  B$_N$A  \\
NGC~7314  & 22 35 46.0 & $-$26 03 02 & 3.0   &  1.2   &   31      & 0&.034   &    2&.74   &  B$_N$A  \\
          &  &    &     &          &      &       &         &     \\
\hline
\end{tabular}
\end{center}
\end{table*}

Many of those sources lying north of $\delta \sim -30\deg$ have been
observed in previous studies. The data available in the literature
will be included in our discussion in $\S$\ref{discussion}.

Thoughout the paper we adopt a Hubble constant
H$_\circ$=75 \kms\ Mpc$^{-1}$.

\subsection{VLA Observations}

VLA snapshot observations were obtained for eight of the survey
Seyferts in the declination range $-30\degr < \delta < 0\degr$.
The VLA observations are summarized in Table~1. The observations were
carried out using the standard 6~cm (4.9~GHz) continuum mode, that is,
with two 50~MHz-wide channels at bandwidth-separated frequencies
(4.835 \& 4.885 GHz).
Three sources were observed using the B-array (8 Apr 1993); the
remaining five were observed using the B$_N$\footnote{The
B$_N$ configuration comprises the east-west portion of the B-array and
and extended northern arm; this configuration provides roughly B-array
resolution for southern sources.} array (1 Feb 1993).

Data reduction followed standard procedures using the NRAO software
package AIPS. The calibration sequence includes: (1) calibrating the
flux scale against observations of the flux standards 3C~48 \&
3C~286, adopting the standard scaling of \cite{baa77} (1977) and
adjusting for calibrator variability; (2) bootstrapping this flux
scale to observations of the phase calibrators; (3) performing
station-based phase and amplitude calibrations on the phase
calibrators; and (4) applying the phase calibrator solutions to the
target sources.  Four iterations of phase-only self-calibration were
then applied to each source.  Solution intervals were typically 5--10
minutes for each iteration of self-calibration.

\subsection{ATCA observations}

The ATCA observing parameters are summarized in Table~2. Twenty-two
objects in the most southern part of this sample ($\delta <
-30\arcdeg$) were observed at 3.5~cm (8.6~GHz) with the ATCA in 6~km
configuration. The only overlap with the VLA observations is
NGC~3393. 

\begin{table*}
\begin{center}
{\bf Table 2: ATCA Observations (3.5~cm)}

\medskip

\def\mccc{\multicolumn{3}{c}}

\begin{tabular}{lrrrrrrrr} \hline
Object & ${\alpha_{2000}}$ & ${\delta_{2000}}$  & \mccc{Resolution}  & rms  &
Peak   \\ 
      &   & & arcsec & arcsec & degrees  & mJy beam$^{-1}$ & mJy beam$^{-1}$  \\ \hline  
TOL~0109-383  &  01 11 27.8 &  --38 04 58 &   2.58 & 0.83 &  46.0 & 0.14  & 10.43 \\
NGC~1365      &  03 33 35.5 &  --36 08 22 &   1.61 & 0.77 & 5.4   & 0.11  & 4.06  \\
NGC~1386      &  03 36 46.3 &  --35 59 58 &   1.68 & 0.75 & --0.6  & 0.13 &  9.88     \\  
NGC~1566      &  04 20 00.3 &  --54 56 18 &   1.29 & 0.75 &  17.5  & 0.17  &  5.65   \\
ESO~362-G18   &  05 19 35.5 &  --32 39 30 &   1.89 & 0.75 & --5.7  & 0.14  &  2.40   \\
IC~2560       &  10 16 19.2 &  --33 33 53 &   1.76 & 0.94 &  --19.3 & 0.26 &   3.86  \\
NGC~3281      &  10 31 52.1 &  --34 51 15 &   1.67 & 0.80 & --16.3 & 0.20 &  16.06   \\
NGC~3393      &  10 48 24.0 &  --25 09 40 &   2.29 & 0.76 & --0.5  & 0.17  & 7.33    \\
NGC~3783      &  11 39 01.7 &  --37 44 19 &   1.59 & 0.74 & --10.5 & 0.24 &  8.15  \\
NGC~4507      &  12 35 37.0 &  --39 54 31 &   1.45 & 0.83 & 25.0  & 0.16  &  3.18   \\
TOL~1238-364  &  12 40 52.8 &  --36 45 21 &   2.42 & 0.88 &  50.0  & 0.16  & 10.29   \\
MCG-6-30-15   &  13 35 53.7 &  --34 17 45 &   1.66 & 0.79 & 1.2   & 0.16  &  0.62  \\
NGC~5643      &  14 32 41.3 &  --44 10 24 &   1.50 & 0.71 & --1.7   & 0.10  &  4.36   \\
ESO~137-G34   &  16 35 14.2 &  --58 04 41 &   1.32 & 0.70 & 32.6  & 0.12  &  2.02   \\
ESO~138-G01   &  16 51 20.5 &  --59 14 11 &   1.42 & 0.71 & 45.9  & 0.14  &  4.34   \\
NGC~6221      &  16 52 46.6 &  --59 12 59 &   1.29 & 0.72 &  27.6  & 0.16  & 2.96   \\
NGC~6300      &  17 16 59.2 &  --62 49 11 &   1.48 & 0.72 &  28.3  & 0.14  &  2.03   \\
IC~5063       &  20 52 02.8 &  --57 04 14 &   1.05 & 0.81 & 51.8  & 0.15  & 160.01  \\ 
IC~5201       &  22 20 57.3 &  --46 02 04 &   1.32 & 0.82 &  3.0   & 0.15  &  $<$0.45  \\
NGC~7496      &  23 09 46.9 &  --43 25 42 &   1.94 & 0.67 &  16.1  & 0.12 &  4.83  \\
NGC~7582      &  23 18 23.1 &  --42 22 11 &   1.29 & 0.78 &  2.7  & 0.10  & 6.22   \\
NGC~7590      &  23 18 55.0 &  --42 14 16 &   1.39 & 0.88 &  7.2   & 0.30  & $<$0.90  \\
\hline
\end{tabular}
\end{center}
\end{table*}

The observations occurred during July, 1995.  Data were taken simultaneously at
8.256 and 8.896 GHz, each through a bandwidth of 128 MHz.  These dual-frequency
observations improved the coverage of the $(u,v)$-plane and the sensitivity. 
Use of dual-frequency synthesis substantially improves the image quality
generated by data from sparse arrays. 

We used the {\sc MIRIAD} package (\cite{sau95}) for data
reduction. The flux density scale was calibrated against observations
of PKS~1934$-$638, assumed to be 2.84~Jy at 8.6 GHz according to the
latest analysis by Reynolds (1996).  Each source was observed in 8
scans with a duration of roughly 20 minutes each. The scans were
spread throughout a 12 hour observing period in order to optimize the
$(u,v)$ coverage within the available integration time. Sources with
particularly weak or interesting structure were repeated in order to
improve the signal-to-noise and image quality.

\section {Results}

Table~3 summarizes our measurements for the 29 observed Seyferts.  The listed
radio morphologies follow the convention of \cite{ulv89}.  The angular and
linear size of the radio emission, measured for those Seyferts with
well-resolved radio structure, were measured to the outermost significant
surface brightness contour; upper limits are given for unresolved Seyferts. 
The radio maps of the resolved objects are presented in Figures 1 to 4 for the
objects observed with the VLA and in Figures 5 to 15 for the objects observed
with ATCA.  To emphasize ring or shell structures that can be lost on contour
maps, we present all maps  as contours on top of greyscale.

\begin{figure}  
\centerline{\psfig{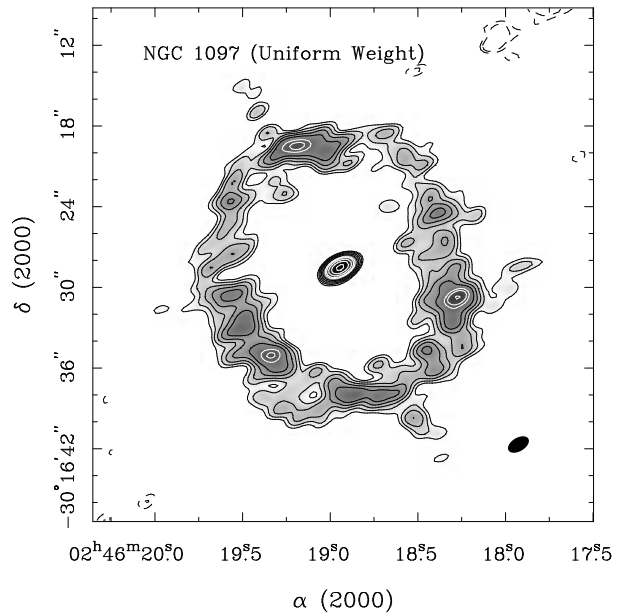}}
\caption{NGC~1097 (uniform weighting)  Contours: 1.0 mJy beam$^{-1}$ $\times$
-0.16, -0.22, 0.16, 0.22, 0.31, 0.44, 0.62, 0.97, 1.21, 1.70, 2.38, 3.33}
\end{figure}

\begin{figure}  
\centerline{\psfig{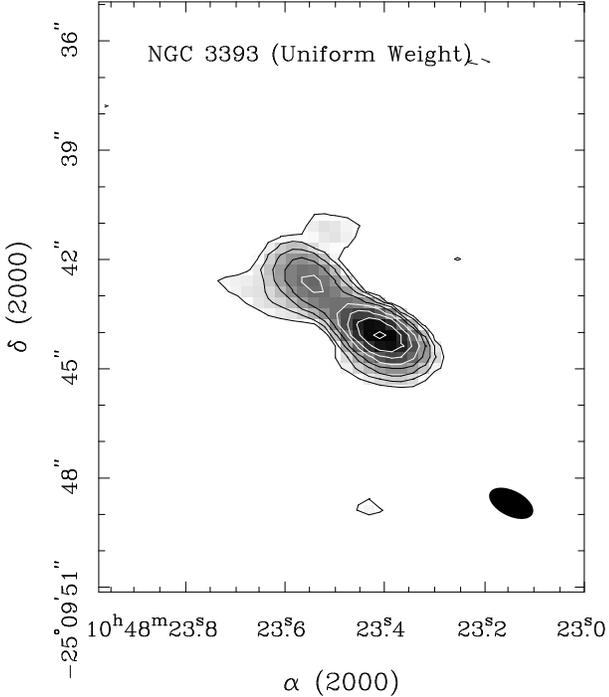}}
\caption{NGC~3393. Contours: 1.0 mJy beam$^{-1}$ $\times$
-0.20, -0.36, 0.20, 0.36, 0.66, 1.20, 2.19, 4.02, 7.36, 13.5}
\end{figure}

\begin{figure}  
\centerline{\psfig{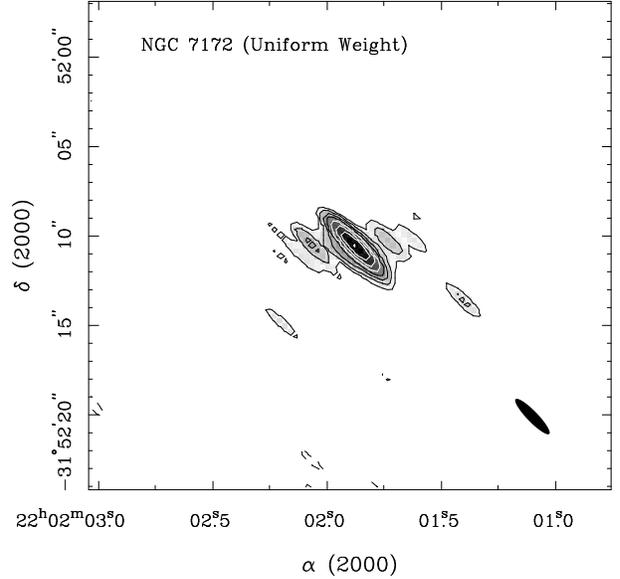}}
\caption{NGC~7172. Contours:
1.0 mJy beam$^{-1}$ $\times$
-0.16, -0.25, 0.16, 0.25, 0.38, 0.57, 0.87, 1.32, 2.00, 3.04}
\end{figure}

\begin{figure}  
\centerline{\psfig{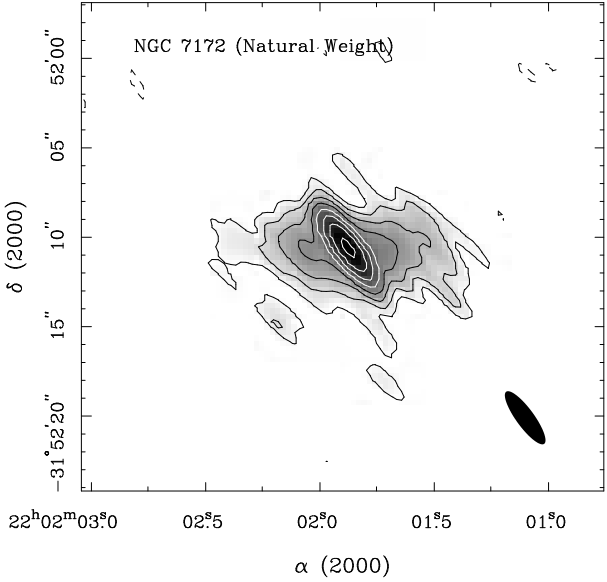}}
\caption{NGC~7172. Contours: 1.0 mJy beam$^{-1}$ $\times$
-0.16, -0.25, 0.16, 0.25, 0.38, 0.58, 0.89, 1.38, 2.12, 3.26}
\end{figure}

\begin{table*}
\begin{center}
{\bf Table 3.} The Radio Parameters \\
 from VLA \& ATCA observations

\medskip

\begin{tabular}{lrrrrrrr} \hline
Object  &      {\em cz}  & Type & ${\rm S_{6cm}}$ & log P$_{6cm}$ & Radio &
LAS & LLS     \\ 
        &        km/s    &   Sey   & mJy & W Hz$^{-1}$ & Morp. & arcsec
& kpc    \\ \hline  
NGC~1097             & 1276 &   1* &   62.3 & 21.33 & U+R &   16.0 &   1.32   \\
NGC~1097~nucleus     & ...  &   1* &    3.8 & 20.12 & U   &$<$ 0.2 &$<$0.02    \\    
NGC~1097~ring        & ...  &   1* &   58.5 & 21.30 & R   &   16.0 &   1.32   \\
NGC~1320              & 2716 &    2 &    2.0 & 20.51 & U   &$<$ 0.8 &$<$0.14  \\
NGC~3393             & 3710 &    2 &   23.1 & 21.83 & L   &    2.7 &   0.65  \\
NGC~3393~SW~peak     & ...  &    2 &   16.0 & 21.67 & U   &$<$ 0.4 &$<$0.10    \\
NGC~3393~NE~peak     & ...  &    2 &    5.0 & 21.16 & S   &    1.3 &   0.31    \\
NGC~3660             & 3300 &    2 &    0.5 & 20.06 & U   &$<$ 1.6 &$<$0.34  \\
NGC~4968             & 3007 &    2 &   13.1 & 21.40 & U   &$<$ 0.7 &$<$0.14   \\
NGC~5427             & 2565 &    2 &    2.5 & 20.54 & U   &$<$ 0.5 &$<$0.08   \\
NGC~7172             & 2527 &    2 &   11.7 & 21.20 & L   &    6.4 &   1.05  \\
NGC~7314             & 1430 &  1.9 &    2.7 & 20.07 & U   &$<$ 0.5 &$<$0.05   \\
\hline
Object  &      {\em cz}   & Type & ${\rm S_{3cm}}$ & log P$_{3cm}$ & Radio & LAS & LLS   \\ 
        &        km/s    &   Sey   & mJy & W Hz$^{-1}$ & Morp & arcsec & kpc    \\ \hline  
TOL~0109-383   &      3496  & 2    &  13.5   & 21.54 & S?     &  3 & 0.68     \\
NGC~1365       &      1662  & 2    &  98.2    & 21.76 & R+L   &   20   &   2.15        \\
NGC~1365~``jet''  &    ...  & ...      &   ...  & ...   & L     &    5   &   0.54       \\
NGC~1386       &       924  & 2     & 12.7   & 20.36 & U     &  $<$1.5 & $<$0.09         \\
NGC~1566       &      1496  & 1     & 8.0    & 20.58 & S?    &     3   &  0.29         \\
ESO~362-G18    &      3777  & 1     & 3.0    & 20.95 & U     & $<$1.5  & $<$0.37    \\
IC~2560        &      2873  & 2     & 6.2    & 21.03 & U     & $<$1.5  & $<$0.28     \\
NGC~3281       &      3460  & 2     & 18.1   & 21.66 & S     &   2.0   &  0.45      \\
NGC~3393       &      3710  & 2     & 13.3   & 21.59 & L     &  3.0    &  0.72          \\
NGC~3393~SW~peak  &    ...  & ...     &  9.3  & 21.43 & ...   & ...  &  ...       \\
NGC~3393~NE~peak  &    ...  & ...    &  3.3   & 20.98 & ...   &  ...   &  ...          \\
NGC~3783       &      3208  & 1     & 10.03   & 21.34 & U     & $<$1.5 & $<$0.31   \\
 NGC~4507       &      3957  & 2     & 5.57   & 21.27 & S     &   3     &    0.77       \\
TOL~1238-364   &      3292  & 2     & 10.58   & 21.39 & U     & $<$2.0 & $<$0.43      \\
MCG-6-30-15    &      2248  & 1     & 1.12    & 20.08 & U     & $<$1.5 & $<$0.22  \\
NGC~5643       &      1199  & 2     & 5.68    & 20.24 & U     & $<$1.5 & $<$0.12     \\
ESO~137-G34    &      2620  & 2     & 9.84    & 21.16 & L     &   4.0    & 0.68    \\
ESO~138-G01    &      2740  & 2     & 7.79    & 21.09 & S     &   3.0    & 0.53  \\
NGC~6221       &      1482  & 2?    & 14.91   & 20.84 & L     &   5.0    & 0.48      \\
NGC~6300       &      1110  & 2     & 2.70    & 19.85 & S     &   3.5    & 0.25   \\
IC~5063        &      3402  & 2     & 229.09  & 22.75 & L     &   4.2    & 0.92      \\
IC~5201        &       915  & 2     & $<$0.45 & $<$18.90 & ... &  ... & $<0.08$    \\
NGC~7496       &      1649  & 2     & 6.52    & 20.57  &  U   & $<$1.5   & $<$0.16   \\
NGC~7582      &      1575  & 2     & 49.29   &  21.41  & D   & 6      & 0.61     \\
NGC~7590       &      1596  & 2     & $<$0.9  & $<$19.69 & ... & ... & $<0.14$     \\ 
\hline
\hline

\end{tabular}

\end{center}

\begin{center}
\begin{tabular}{l} 
{\bf Radio Morphology:} L = linear; R = ring; D = diffuse; A = ambiguous; 
S = slightly resolved; U = unresolved. \\
$^*$ NGC~1097 is a LINER that periodically shows Sy1-like broad lines 
(Storchi-Bergmann, Baldwin, \& Wilson 1993).
\end{tabular}
\end{center}
\end{table*}

Most of the resolved sources bear a linear radio structure as is
commonly observed in Seyfert nuclei. In NGC~1097 and NGC~1365 we
resolve a central, unresolved radio source, apparently related to the
AGN, and a ring associated with star forming regions (compare with
\cite{hum87} 1987 and \cite{san95} 1995, respectively).  NGC~7582
displays diffuse radio structure.  Only two objects of the observed
(IC~5201 and NGC~7590) were undetected; both are part of the group
observed at 3 cm with ATCA.

Previous observations for 9 of the 29 objects have been reported in
the literature by various authors.  Where possible we have used these
data in combination with our new observations to estimate a radio
spectral index, $\alpha$, defined as $S\propto \nu^{\alpha}$. We
caution that there are inestimable uncertainties in the spectral
indices for extended sources owing to differences in the $(u,v)$
coverage between our observations and those in the literature.
Nevertheless, the indices are accurate for unresolved sources and
otherwise provide a sense of the spectral shape for the extended
sources.

In the next section we discuss each source individually and also summarize
relevant radio and optical observations available in the literature.  Where
possible, the flux at 20 cm (1.4 GHz), and the Parkes-Tidbinbilla
interferometer (PTI) flux measurements at 13 cm (2.3 GHz) taken from
literature, are listed.  The PTI observations have an effective resolution of
$\sim 0$\farcs1, much higher that the data presented in this paper.  The main
information we can derive from PTI observations is whether a compact radio
structure ($<0$\farcs1) is present.  Where possible we compute spectral indices
between 20 and 6 cm ($\alpha_{6}^{20}$) and between 6 and 3 cm
($\alpha_{3}^{6}$), and compare with previously reported values. 

Our snapshot observations are insensitive to structures larger than 15--18
arcsec, or around 4~kpc at the redshift limit of this survey. The
observations therefore resolve out diffuse emission from galaxy disks, but
we remain sensitive to radio emission from compact star-forming regions.
An interesting result of this survey is that star-forming rings are not
common. Only NGC~1097 and NGC~1365 display starburst rings, similar to the
one observed in NGC~1068, for example, even though our observations would
have detected and resolved such rings out to the redshift limit of the
survey. Otherwise the radio structures resolved by this survey resemble
the classic linear radio structures commonly observed in Seyfert nuclei,
and we assume that these structures are jets powered by the AGN. Of
course, we are unable to distinguish jets and compact nuclear starbursts
in unresolved sources, corresponding to linear scales $\la 500$~pc at the
redshift limit of the survey.

\subsection {Notes on individual sources}

\noindent {\sl TOL~0109-383 (NGC~424)}: 
This object has been observed by \cite{ulv89} at 20 and 6 cm. With
$\sim1$\arcsec\ resolution they found a slightly extended source with
flux of 22.3 mJy at 20 cm, and 14.9 mJy at 6 cm. In our data we also
find a structure that is slightly extended to the east (Fig.5), although the
very elongated beam of our observations make this very uncertain. \cite{ulv89}
found that the dominant component has a flat spectrum with a spectral
index between 20 and 6 cm of $\alpha^6_{20} = -0.17$. Our 3 cm
measurement shows that the spectrum remains flat between 6 and 3 cm,
$\alpha^3_6=-0.21$.\\

\noindent {\sl NGC~1097}: 
Originally classified as a LINER by the optical emission-line
spectrum, the recent appearance of broad Balmer-line emission and a
featureless blue continuum implies that it has a Seyfert 1 nucleus
(Storchi-Bergmann, Baldwin, \& Wilson 1993).  The radio structure
(Fig.\ 1) comprises an unresolved point source and the well-known
star-forming ring.  A detailed radio study of this object was
performed by \cite{hum87} (1987) and shows an overall steep spectrum
($\alpha$ between $-0.6$ and $-0.8$) and an inverted spectrum ($\alpha
= 1.0$) for the nuclear component.  The flux of the nucleus measured
from our data is very similar to that found by \cite{hum87} (1987).
An upper limit to the core flux ($S_{13cm}<5$ mJy) has been obtained
from PTI observations by \cite{sad95} (1995, hereafter S95).\\

\noindent {\sl NGC~1365}: 
This is a well known southern barred galaxy.  It was observed in the
radio continuum (20, 6 and 2 cm) by \cite{san95} (1995, hereafter
SJL95).  As in SJL95, our map (Fig.\ 6) shows a ring of emission with
angular dimensions $8\arcsec \times 20\arcsec$.  This ring-like
emission is similar to that found in NGC~1097.  SJL95 also identify
the existence of a jet-like structure originating from the nucleus and
about $5\arcsec$ long in position angle (PA) $125\arcdeg$ (i.e., aligned
with the minor axis of the galaxy).  We observe a similar structure in
our map.  The detailed study of the spectral data (SJL95) indicate
that both the jet and the nucleus have a steep spectral index.  This
jet-like feature appears to be aligned with the axis of the ionized
gas.  NGC~1365 has also been observed with the PTI at 13 cm ($\sim$
$0.1\arcsec$ resolution) by \cite{roy94} (hereafter R94) and S95
in which a 4 mJy component was detected.  This object has been
extensively studied in \HI\ by \cite{jor95} (1995).\\

\noindent {\sl NGC~1320 (Mrk~607)}: 
This object is unresolved in our VLA observations. PTI observations
place an upper limit of $S < 4$ mJy on the compact flux density at
13cm (R94).\\

\noindent {\sl NGC~1386}: 
\cite{ulv84b} observed this object at 20
and 6~cm.  In their observations, NGC~1386 is barely resolved with an
extension to the southwest (PA=$-125\arcdeg$), and it is unresolved in
our 3.5~cm map.  \cite{ulv84b} give a flux density of 13 mJy and 23.0 mJy at 6
and 20 cm, respectively, and a spectral index of $\alpha^6_{20} =
-0.47$.  The spectral index between 6 and 3cm is very flat,
$\alpha^3_6 = -0.05$.  PTI observations find a flux density of 4~mJy
at 13~cm (R94, S95).\\

\noindent {\sl NGC~1566}: 
A possible faint blob of radio emission is detected 3\arcsec\ north of
central peak (Fig.~7).  Observations with the PTI at 13cm (R94 and S95) give
a 5 mJy flux.\\

\noindent {\sl ESO~362--G18} : This object is not resolved by our
observations.  PTI observations give an upper limit of $S < 4$ mJy on
the flux density at 13 cm (R94). ESO~362--G18 has been studied in
\Oiii and H$\alpha$ by Mulchaey \etal\ (1996).\\

\begin{figure}  
\centerline{\psfig{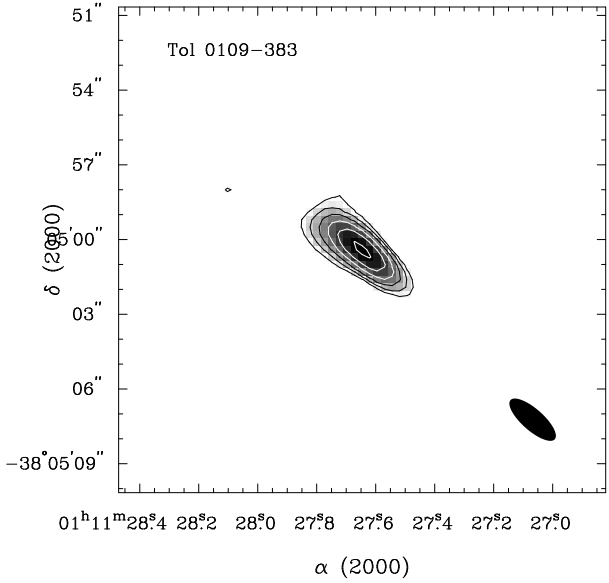}}
\caption{TOL~0109--383. Contours: 
--1.0, --0.6, 0.6, 1, 1.7, 3, 5, 9  mJy  beam$^{-1}$}
\end{figure}

\begin{figure}  
\centerline{\psfig{figure=ds1636f6.eps,angle=-90,width=8cm}}
\caption{NGC~1365. Contours: 
--0.6, --0.4, 0.4, 0.6, 0.9, 1.5, 2.3, 3.6  mJy  beam$^{-1}$ }
\end{figure}

\begin{figure}  
\centerline{\psfig{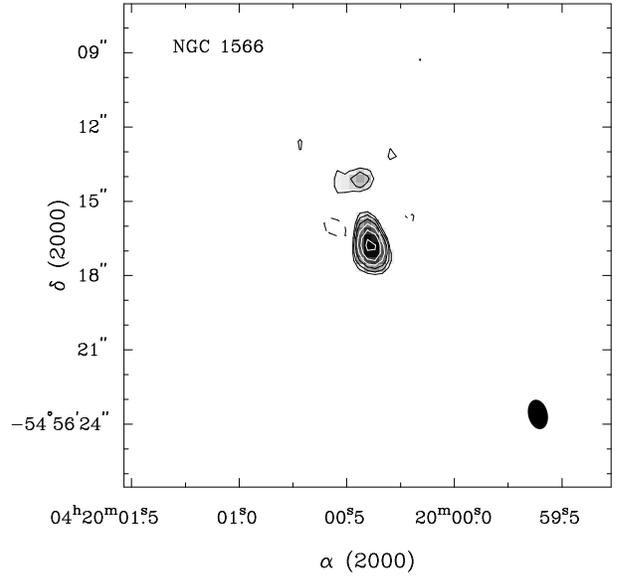}}
\caption{NGC~1566. Contours: --0.9, --0.6, 0.6, 0.9, 1.4, 2.2, 3.4, 5.3  mJy  beam$^{-1}$}
\end{figure}

\begin{figure}  
\centerline{\psfig{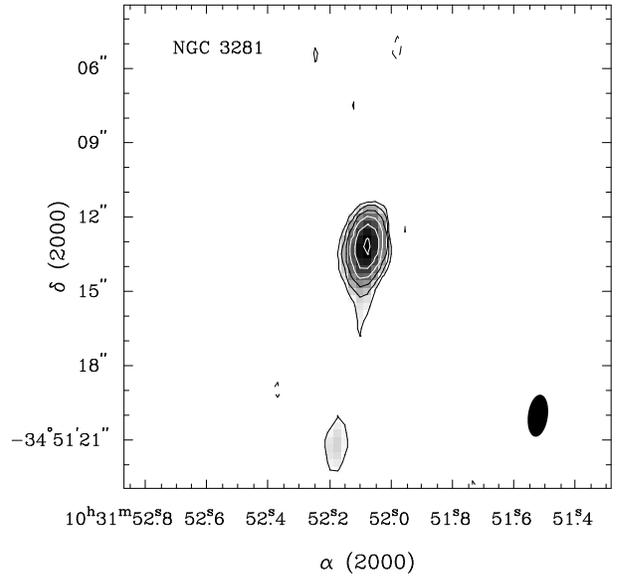}}
\caption{NGC~3281. Contours: --0.9, --0.5, 0.5, 0.9, 1.9, 3.7, 7.2, 14.2 
mJy  beam$^{-1}$}
\end{figure}

\begin{figure}  
\centerline{\psfig{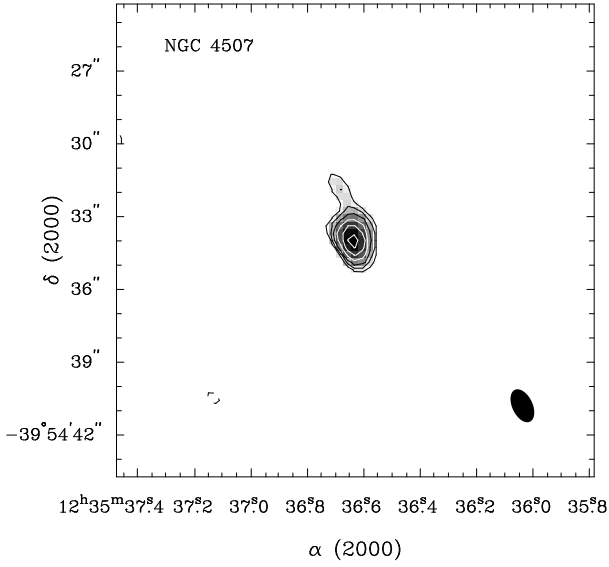}}
\caption{NGC~4507. Contours: --0.7, --0.5, 0.5, 0.7 1, 1.4, 2, 3  mJy 
beam$^{-1}$}
\end{figure}

\begin{figure}  
\centerline{\psfig{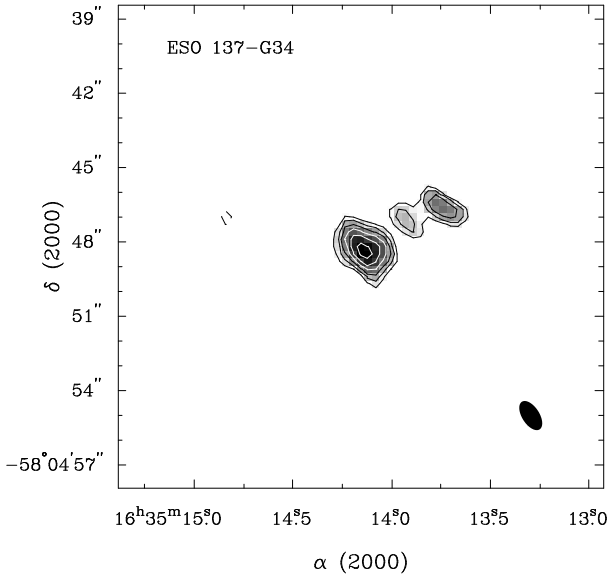}}
\caption{ESO~137--G34 Contours:
--0.5, --0.4, 0.4, 0.5, 0.7, 1, 1.3, 1.8 mJy  beam$^{-1}$}

\end{figure}

\begin{figure}  
\centerline{\psfig{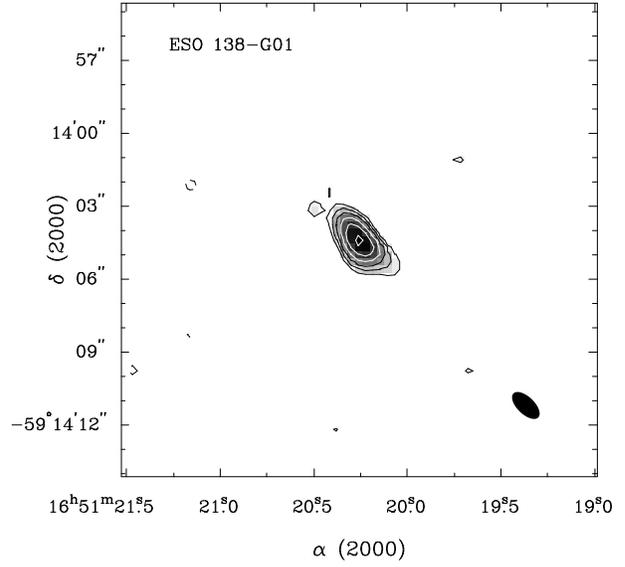}}
\caption{ESO138-G01 Contours:
--0.7, --0.4, 0.4, 0.7, 1.1, 1.6, 2.5, 3.9 mJy  beam$^{-1}$}

\end{figure}

\begin{figure}  
\centerline{\psfig{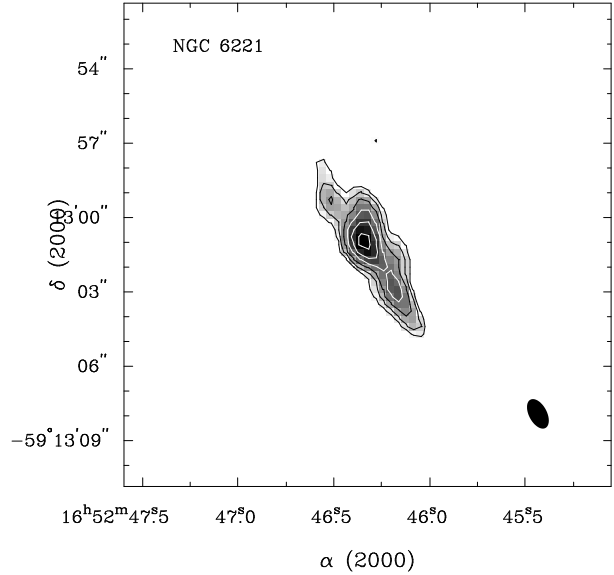}}
\caption{NGC~6221 Contours:
--0.6, --0.4, 0.4, 0.6, 0.9, 1.3, 1.8, 2.7 mJy  beam$^{-1}$}

\end{figure}

\begin{figure}  
\centerline{\psfig{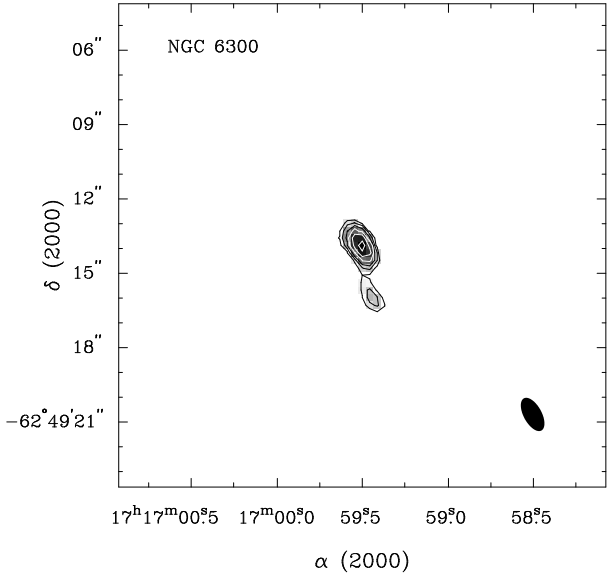}}
\caption{NGC~6300 Contours:
--0.7, --0.5, 0.5, 0.7, 0.9, 1.1, 1.4, 1.8 mJy  beam$^{-1}$}

\end{figure}

\begin{figure}  
\centerline{\psfig{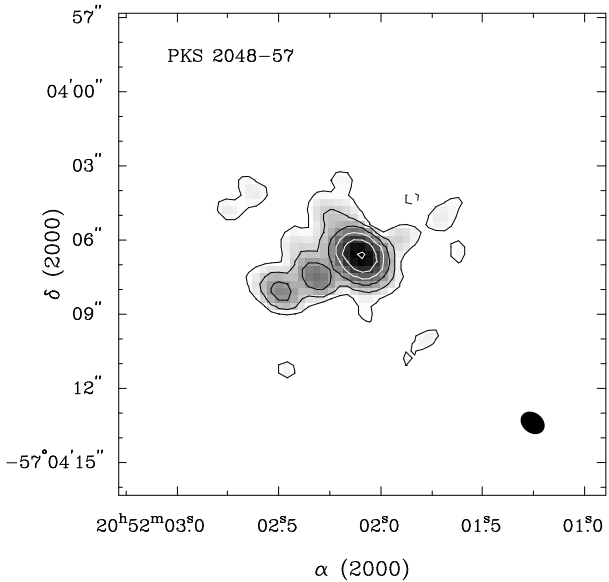}}
\caption{IC~5063 Contours:
--1.8, --0.6, 0.6, 1.8, 5.5, 16.3, 48.4, 144.1 mJy  beam$^{-1}$}

\end{figure}

\begin{figure}  
\centerline{\psfig{figure=ds1636f15.eps,angle=-90,width=8cm}}
\caption{NGC~7582 Contours:
--0.6, --0.4m 0.4, 0.6, 0.9, 1.5, 2.3, 3.6, 5.6 mJy beam$^{-1}$}

\end{figure}

\noindent {\sl NGC~3281}: This object is possibly barely resolved (along the
north-south direction) in our data (Fig.~8), and was not resolved by \cite{ulv89}.
\cite{ulv89} find a flux of 61.2 mJy at 20 cm and 26.7 mJy at 6 cm which gives
a steep spectral index of $\alpha^{6}_{20} =-0.65$.  Our observations
show that the spectral index remains steep out to 3 cm, ($\alpha^3_6 =
-0.83$).\\

\noindent {\sl NGC~3393}: This object was observed with both the VLA and ATCA. 
The two radio maps are very similar and show a double structure.  Here
we show only the VLA map (Fig.\ 2), which has a slightly better beam
shape. From the two frequencies we derive a spectral index of
$\alpha^3_6 =-0.71$ for the NE side and $-0.93$ for the SW side.  The
NLR is elongated along a similar position angle as the radio --
PA$_{\rm NLR} \approx$ PA$_{\rm radio} = 45\arcdeg$ -- with a close
correspondence to the radio morphology.  This is very clear in the HST
image (Pogge 1997) which shows a spectacular S-shape morphology.  A
core of 16 mJy was detected with the PTI at 13 cm (R94).\\

\noindent {\sl NGC~3660}: This object is unresolved in our observations.  The
source was previously detected by the Effelsberg 100m telescope (Kollatschny et
al.\ 1983).  The total flux measured in our observation (0.5 mJy) is much
smaller than that reported by Kollatschny et al.\ (11 mJy).  The reason for
this discrepancy is likely to be the presence of other two bright unrelated
sources in the field.  These sources could not be separated from NGC~3660 by
the 100m telescope but they are in our VLA data. \\

\noindent {\sl NGC~3783}: 
The object is unresolved in our 8 GHz observations and was also
reported unresolved by \cite{ulv84b} who give a flux density of 13 mJy at 6 cm.
A spectral index of $\alpha^3_6 = -0.55$ is derived from \cite{ulv84b} and our
flux measurements.  An upper limit ($S < 5$ mJy) was obtained from the
PTI observations at 13 cm (R94).\\

\noindent {\sl NGC~4507}: In our 3 cm data, this object is slightly extended
along PA$\sim 10\arcdeg$ (Fig.~9).  This face-on galaxy shows an high
excitation NLR with a fairly symmetric edge-brightened bicone shape.\\

\noindent {\sl TOL~1238--364 (IC~3639)}: 
In our 3 cm map the source is unresolved.  TOL~1238--364 object was observed
at 20 and 6 cm by \cite{ulv89}.  At 20 cm the source appears to have a diffuse
emission around a relatively strong core.  At 6 cm only the core of
the source was detected.  \cite{ulv89} find a 6 cm flux of 13.6 mJy which
combined with our data gives a spectral index of $\alpha^3_6 = -0.53$.
A core of 13 mJy was detected with the PTI at 13 cm (R94).\\

\noindent {\sl NGC~4968}: 
This object is unresolved in our observations.  A core of 10 mJy has
been detected with the PTI at 13 cm (R94).\\

\noindent {\sl MCG-6-30-15}:
This is an unresolved and barely detected object.  \cite{ulv84b} reported a flux
density of 1 mJy at 6 cm and 1.7 at 20 cm, giving a spectral index of
$\alpha^{20}_6 = -0.44$.  The flux density from our observations give an
inverted spectral index at high frequencies $\alpha^3_6 = 0.24$.  The upper
limit to the flux with the PTI is $S < 4$ mJy at 13 cm (R94). \\

\noindent {\sl ESO~137-G34}: This is a low-surface brightness spiral
classified as Seyfert 2.  The radio morphology (Fig. 10) consists of
three knots aligned along PA$_{\rm radio} \sim -40\arcdeg$.  The
ionized gas has an S-shape morphology within a larger scale bi-cone.
The radio emission is coincident with the inner (linear) part of the
line-emitting gas.  \\

\noindent {\sl ESO~138-G01}: This is a Seyfert 2 nucleus residing in early
type host galaxy.  The radio emission (Fig.  11) is possibly elongated in
PA$_{\rm radio} \sim 35\arcdeg$.  The NLR has a jet-like morphology elongated
PA$_{\rm NLR}\sim -40\arcdeg$ The radio emission is also misaligned by $\sim
50\arcdeg$ with respect to the major axis of the galaxy.  \\

\noindent {\sl NGC~6221}: We resolve a radio triple (see Fig.12) aligned along
PA$_{\rm radio} \sim 40\arcdeg$.  The upper limit to the core flux
with the PTI at 13 cm is $S < 2$ mJy (S94).\\

\noindent {\sl NGC~6300}: The radio morphology (Fig. 13) of this object is
slightly elongated.\\

\noindent {\sl IC~5063 (PKS~2048-57)}: This is an early-type galaxy hosting a
Seyfert 2 (Colina, Sparks \& Macchetto 1991).  Its radio luminosity is
nearly 100 times larger than typical values for nearby Seyferts.  Our
ATCA radio map (Fig.~14, see also Morganti, Oosterloo \& Tsvetanov
1998) resolves a linear radio structure comprising three compact
features aligned along PA$_{\rm radio} = 305\arcdeg$.  The spectral
index is steep: $\alpha^{20}_6 = -1.1$ (Danziger, Goss \& Wellington
1981).  The optical line-emitting gas traces an {\bf X}-shape
(Morganti et al. 1998), and the radio emission coincides with the
inner part of the line emitting region.  Recent \HI\ observations with
ATCA and VLBI reveal a well defined \HI\ disk aligned with a
morphologically similar system of dust lanes (Morganti, Oosterloo \&
Tsvetanov 1998, Oosterloo et al.  1998).  PTI observations measure a
flux of 140 mJy at 13 cm and 10 mJy at 3 cm \cite{sle94} (1994).\\ 

\noindent {\sl NGC~7172}:
This object has a linear radio structure (Figs. 3 and 4) which is
elongated roughly east-west (PA$_{\rm radio} = 90\arcdeg$). A 13~cm core
flux of 3 mJy has been detected with the PTI (R94).\\

\noindent {\sl IC~5201}: 
This source is undetected in our observations. An upper limit $S < 5$
mJy has been obtained from PTI observations at 13 cm (S94).\\

\noindent {\sl NGC~7496}:  
This source is unresolved by our observations. A core flux of 7 mJy
has been detected with the PTI at 13 cm (R94).\\

\noindent {\sl NGC~7582}: 
\cite{ulv84b} detected a weak core surrounded by an extended emission.  They
measure 69 mJy at 6 cm and 166 mJy at 20 cm, giving a spectral index
of $-0.73$.  We find (Fig. 15) that the spectral index remains steep
out to 3 cm with $\alpha^3_6 = -0.72$. For comparison, the NLR is a
good example of an edge-brightened, wide angle bi-cone with a cone
axis projected along PA$_{\rm e.l.} = 55\arcdeg$.  In addition, the
host galaxy has a prominent bar along PA$_{\rm bar} = 155\arcdeg$.  An
upper limit to the flux density of $S < 6$ mJy has been set by the PTI
observations (S95, R94).\\

\noindent {\sl NGC~7590}: 
This object is not detected in our data; the upper limit to the flux
density from PTI is $S < 3$ mJy at 13 cm (R94).\\

\begin{table*}
\begin{center}

{\bf Table 4.} Radio Parameters from the Literature

\medskip

\begin{tabular}{lrrrrrrr} \hline
Object    & {\em cz} & Sey & $S_{\rm 6cm}$ & log $P_{\rm 6cm}$ & Radio    &  Size &  Ref. \\ 
          &  km/s    &Type & mJy           & W Hz$^{-1}$       & Morph.   &  kpc  & \\ \hline  
NGC~2639    &  3336 &  1 &   54.5 & 22.11 &  L    & 0.41 & 4 \\ 
NGC~3227    &  1157 &  1 &   34.0 & 20.99 &  S    & 0.30 & 3 \\ 
NGC~3516    &  2649 &  1 &    4.3 & 20.81 &  U    & $<$0.05 & 3 \\ 
NGC~3786    &  2678 &  1 &    3.4 & 20.72 & S     & 0.28 & 3\\ 
NGC~4051    &   725 &  1 &    6.0 & 19.83 & L-D?  & 0.57 & 3 \\ 
NGC~4151    &   995 &  1 &  125.0 & 21.42 & L     & 0.34 & 3, 5 \\ 
NGC~4235    &  2410 &  1 &    5.3 & 20.82 & U     & $<$0.05 & 3 \\ 
NGC~4253    &  3876 &  1 &   19.5 & 21.80 & U     & $<$0.35 & 4\\ 
NGC~4593    &  2698 &  1 &    1.6 & 20.40 & U     & $<$0.04 & 3 \\ 
NGC~5033    &   875 &  1 &    3.3 & 19.73 & U     & $<$0.02 & 4\\ 
NGC~5273    &  1054 &  1 &    0.9 & 19.33 & S     & 0.10 & 3\\ 
NGC~6814    &  1563 &  1 &    2.2 & 20.06 & S     & 0.03 & 3 \\ 
0714-2914   &  1630 &  2 &   28.2 & 21.20 & L     & 0.55 & 4\\ 
0942+0950   &  3897 &  2 & $<$0.4 & $<$20.11 & U     & $<$0.10 & 4\\ 
Mrk~3       &  4050 &  2 &  361.0 & 23.10 & L     & 0.58 & 2, 6\\ 
Mrk~270     &  2700 &  2 &    5.7 & 20.95 & L     & 0.61 & 2\\ 
Mrk~348     &  4540 &  2 &  480.0 & 23.32 & L     & 0.06 & 2, 7\\ 
Mrk~1066    &  3605 &  2 &   35.5 & 21.99 & L     & 0.70 & 4\\ 
NGC~0591    &  4547 &  2 &    7.9 & 21.54 & L     & 0.18 & 4\\ 
NGC~0788    &  4078 &  2 &    1.2 & 20.63 & U     & $<$0.08 & 4\\ 
NGC~1068    &  1136 &  2 & 1090.0 & 22.48 & L     & 1.43 & 1, 8 \\ 
NGC~1358    &  4013 &  2 &    1.2 & 20.62 & S    & 0.10 & 4\\ 
NGC~1667    &  4547 &  2 &    1.0 & 20.64 & S     & 0.12 & 4\\ 
NGC~1685    &  4527 &  2 &    5.3 & 21.37 & S     & 0.12 & 4\\ 
NGC~2273    &  1840 &  2 &   19.0 & 21.14 & L     & 0.24 & 3\\ 
NGC~3081    &  2385 &  2 &    0.9 & 20.04 & U     & $<$0.06 & 3 \\ 
NGC~3982    &  1109 &  2 &    2.2 & 19.76 & U     & $<$0.03 & 4\\ 
NGC~4117    &   871 &  2 & $<$0.6 & $<$18.99 & S     & $<$0.08 & 4\\ 
NGC~4388    &  2524 &  2 &   76.0 & 22.02 & L     & 3.92 & 3 \\ 
NGC~4941    &  1108 &  2 &    4.3 & 20.05 & S     & 0.10 & 3 \\ 
NGC~5135    &  4112 &  2 &   58.8 & 22.33 & A     & 1.97 & 4\\ 
NGC~5347    &  2335 &  2 &    2.2 & 20.41 & U     & $<$0.06 & 4\\ 
NGC~5695    &  4225 &  2 & $<$0.5 & $<$20.28 & U     & $<$0.38 & 4\\ 
NGC~5728    &  2788 &  2 &    4.6 & 20.88 & L?    & 2.34 & 3, 9 \\ 
NGC~5929    &  2561 &  2 &   24.7 & 21.54 & L     & 0.18 & 3, 10\\ 
NGC~6890    &  2419 &  2 &    4.2 & 20.72 & U     & $<$0.06 & 4 \\ 
NGC~7450    &  3191 &  2 &    1.7 & 20.57 & L?    & 0.62 & 3 \\ 
NGC~7672    &  4117 &  2 &    1.0 & 20.56 & L     & 1.86 & 4\\ 
Tol~0074    &  3285 &  2 &   13.6 & 21.50 & D     & $<$0.09 & 4\\ 
MCG-5-23-16 &  2482 &  2 &    6.0 & 20.90 & S     & 0.08 & 3 \\ 
NGC~2110    &  2284 &  2 &  175.0 & 22.29 & L     & 0.87 & 3 \\ 
NGC~2992    &  2311 &  2 &   77.0 & 21.94 & L-D?  & 1.36 & 3, 11 \\ 
NGC~5506    &  1853 &  2 &  160.0 & 22.07 & S     & 0.65 & 3 \\ 
\hline

\end{tabular}
\end{center}

{\bf References:} (1) Wilson \& Ulvestad (1983); (2) Ulvestad \& Wilson 1984a;
(3) Ulvestad \& Wilson 1984b; (4) Ulvestad \& Wilson 1989; (5) Pedlar et al. 
1993; (6) Kukula et al.  1993; (7) Neff \& de Bruyn 1983; (8) Gallimore et al. 
1996; (9) Schommer \etal\ 1988; (10) Su et al.  1996; (11) Wehrle \& Morris
1988

\end{table*}

\section{Discussion}\label{discussion}

The new radio observations presented in this paper cover more than 50\% of our
southern ($\delta<0\arcdeg$) volume limited ($cz<3600$ \kms ) sample of
Seyfert galaxies; adding data from the literature increases the coverage to
almost 85\%.  There remain 9 sources in the southern sample for which radio
information at arcsec resolution is not available yet. 

Our survey largely overlaps with the samples studied by
\cite{ulv84b} and \cite{ulv89}.  We do not, therefore, expect results
completely independent from theirs.  We note, however, that the sample of
Ulvestad \& Wilson contained all Seyferts with $cz < 3600$ km s$^{-1}$ known
at the time of their observations ($\sim 15$ years ago), and was
declination-limited by the VLA horizon restrictions.  Our sample, on the other
hand, covers all Seyferts with reliable classification to date at southern
declinations ($\delta<0\arcdeg$) and increases the distance-limited sample of
\cite{ulv89} by 22\%.  

However, in order to further improve the statistics, we have collected
additional sources from the literature, consisting of known Seyferts with $cz <
4600$ \kms, derived mainly from \cite{ulv89} and references therein.  The
selected sources are listed in Table~4.  For this compilation, we selected only
radio data of similar observing frequency and resolution to our observations
($\sim 0.3$--1\arcsec).  Most of the collected data were obtained at 6 cm.  The
combined sample includes 71 objects of which 17 are Seyfert 1's and 54 Seyfert
2's. 

Here we concentrate on the discussion of the radio characteristics of the
studied sample.  A discussion on the comparison between the radio and optical
properties will be done in a forthcoming paper. 

\subsection{Radio Spectra} 

We derived the spectral indices for most of our sources by combining
our data with previous observations (see $\S$3.1). The exception is
NGC 3393, for which we use the 3 and 6~cm data presented here.  In
three cases (NGC~1097, NGC~1365 and IC~5063), we are unable to
estimate the spectral index owing to complex source structure and poor
matches in $(u,v)$ coverage.  For NGC~1097 and NGC~1365 detailed
studies of the spectral indices are already available (Hummel, van der
Hulst \& Keel 1987; Sandqvist, J\"ors\"ater \& Lindblad 1995
respectively).

In three cases (NGC~7582, NGC~3281 and Tol~0109-383) we find that the spectral
index at high frequencies ($\alpha_{\rm high}$, between 3 and 6~cm) is similar
to the value at lower frequencies ($\alpha_{\rm low}$); that is, the spectra
are steep in the first two cases, and flat for Tol~0109-383.  In two cases
(MGC-6-30-15 and NGC~1386) the spectra appear to flatten with increasing
frequency ($\alpha_{\rm high} > \alpha_{\rm low}$), although the radio source
in MGC-6-30-15 is very weak, and the derived spectral index is uncertain.  New
spectral indices have been derived for NGC~3393, NGC~3783 and Tol~1238-364: in
all these cases the spectral index is steep ($\alpha <-0.5$).  For comparison,
most of the Seyfert galaxies have steep radio spectra, but flat-spectrum cores
are found in a few Seyferts (Wilson 1991). 

Accepting uncertainties in our spectral index measurements due to mismatched
$(u,v)$ coverage in the 3 and 6 cm maps, we measure a median spectral index
steep ($\alpha \sim -0.67$) for our Seyfert nuclei.  This result is in
agreement, with the spectral indices of the inner $0\farcs1$ nuclear regions
of spirals (and Seyferts) as measured by \cite{sad95} (1995) using PTI
observations.  Their investigation of the compact radio cores in spiral and
elliptical galaxies found a median spectral index for spirals (and Seyferts)
of $\alpha=-1.0$, and that spirals usually have steeper core spectra than do
elliptical galaxies (median spectral index $\alpha = +0.27$).  The flat (or
inverted) spectral index of the cores is a typical characteristic of
elliptical galaxies found on all scales in which the nuclear regions have been
observed (from arcsec and sub-arcsec scale, see e.g.  Morganti et al.  1997
and Slee et al.  1994 to VLBI scale).  Thus, our study confirms the result
that the spectral indices of Seyferts are much steeper than in the cores of
ellipticals, and that the spectral index remains close to the same value over
$0\farcs1$ to 1\arcsec\ scales.

In connection with this difference in spectral index, it is worth remembering
that the nuclear regions of Seyfert galaxies appear to have a more complicated
situation than in radio galaxies: following the detailed studies of few well
objects (e.g.  NGC~1068, Gallimore et al.  1996, Roy et al.  1998; NGC~4151
Ulvestad et al.  1998) free-free absorption appears to be relevant in Seyfert
to dim the "real" radio core.  Thus the nuclear emission (and its spectral
index) can be dominated not by the core itself but by bright blobs.  This
would be in agreement with the finding (Sadler et al.  1995) that in
ellipticals most of the radio emission in the central kpc comes from the
parsec-scale core, while in Seyferts this is only a small fraction (10-25\%).

\subsection {Radio Structure}

Table~3 summarizes the classification of the radio morphology for the
newly observed objects, following the scheme of \cite{ulv84b}.  Table~4
provides source classifications for data taken from the literature.
Including Seyferts from the literature, we find 38\% of the sample are
unresolved sources, 23\% have slightly resolved structures and 34\%
have linear structure at arcsecond resolution; the remaining 5\% have
diffuse, amorphous, or ringed structures.

\begin{figure*}
\centerline{\psfig{figure=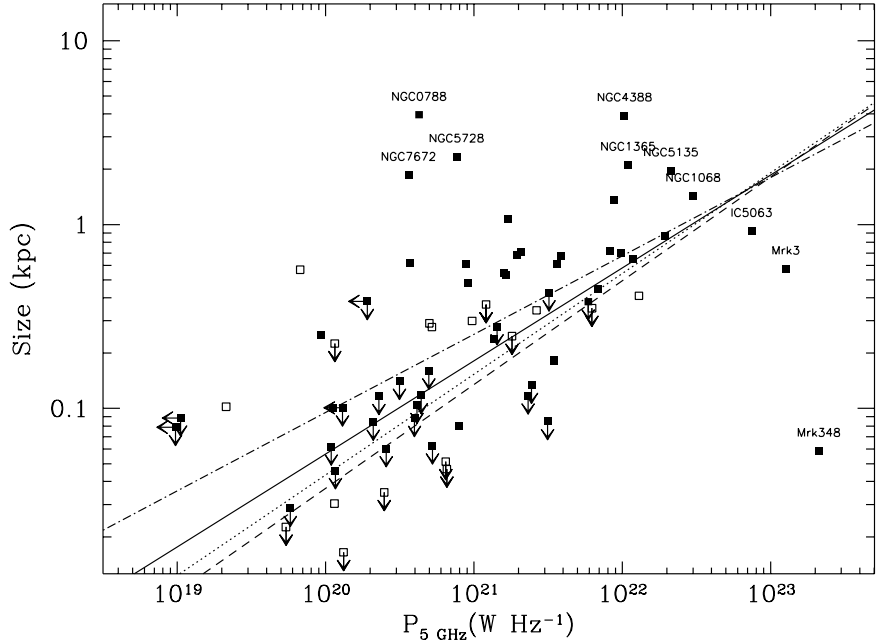,angle=-90,width=12cm}}
\caption{ Plot of the size of the radio emission versus the 
          radio power (at 5 GHz): open squares  represent Seyfert 
          1 and filled squares represent Seyfert 2.}
\end{figure*}

We can compare these numbers with the results from \cite{ulv89} and references
therein (see also the summary in \cite{wil91} 1991). Compared to
\cite{wil91} (1991), we measure a higher fraction of unresolved
sources and a lower fraction of galaxies with diffuse radio emission.
This result may be due to an observational bias; our observations are
not as sensitive to diffuse, extended, steep spectrum emission as are
the 20~cm VLA observations of, e.g.,  Ulvestad \& Wilson (1989).
Otherwise, we measure a similar detection fraction of linear or
slightly resolved structures.

Linear radio sources in Seyfert galaxies generally trace radio outflows,
but it is worth checking that linear radio emission is not associated with
star formation in an edge-on galaxy. Among the linear radio sources
detected in this survey, only the host galaxy NGC 7172 is near edge-on. 
The radio structure aligns with the plane of the host galaxy,
casting doubt on a jet origin; rather, it appears that the linear radio
source of NGC~7172 may be associated plausibly with a nuclear starburst.
Otherwise, the remaining linear radio sources are more likely associated
with an AGN-driven outflow.

To compare the distributions of radio power, we adjusted all of the radio
luminosities to their 6 cm values.  For sources not observed at 6~cm, we
adjusted the luminosities using measured spectral indices where possible.  For
those sources having neither 6~cm measurements or measured spectral indices,
we assumed a spectral index of $\alpha^3_6 = -0.5$.  Many of the low power
sources are unresolved, resulting in an upper limit to the source size.  Since
we are interested in the compact radio emission from jets rather than
extended, low surface brightness emission (which may come from star-forming
regions and starburst-driven superwinds), we also assumed that the three
undetected sources were unresolved (size upper limit) and radio weak (flux
upper limit).  To account appropriately for these limits, we employed survival
analysis techniques from the ASURV package (Feigelson \& Nelson 1985; Isobe,
Feigelson, \& Nelson 1986) as it is implemented in IRAF. 

We confirm a correlation between radio power and size (Fig.~16), originally
discussed by Ulvestad \& Wilson (1984b) and \cite{giu90} (1990).  Both the
power-size and flux-angular size correlations are significant at better than
1\% (probability of no correlation) according to the traditional survival
analysis tests (Kendall's $\tau$, Spearman's $\rho$, and the Cox Proportional
History model).  A summary of log-linear models for the power-size correlation
is provided in Table~5.  Comparing with the results of \cite{ulv84b}, we find
a steeper slope in the correlation owing to a proper treatment of limits at
low radio-powers.  Mrk~348 is an outlier, falling at relatively small size for
its radio power.  The present analysis ignores the $\sim 5$ kiloparsec radio
lobes in this source, which may arise from either a starburst-driven
superwind, or old nuclear-driven ejecta.  Accounting for the extended radio
lobes places Mrk~348 closer to the best-fit line, except that Mrk~348 now
falls somewhat oversized for its luminosity according to the correlation.

\begin{table}
\begin{center}

{\bf Table 5.} Log-linear fit to  Radio Power vs. Size

\medskip

\begin{tabular}{lrrrrrrr} \hline
Method & Intercept & Err & Slope & Err \\
\hline  
Buckley-James & -11.36 &  ...  & 0.51 & 0.09 \\
E-M           & -12.31 & 2.25  & 0.55 & 0.11 \\
Schmitt       & -12.72 & 2.20  & 0.56 & 0.10 \\
\hline
\end{tabular}
\end{center}
\end{table}

\begin{figure}
\centerline{\psfig{figure=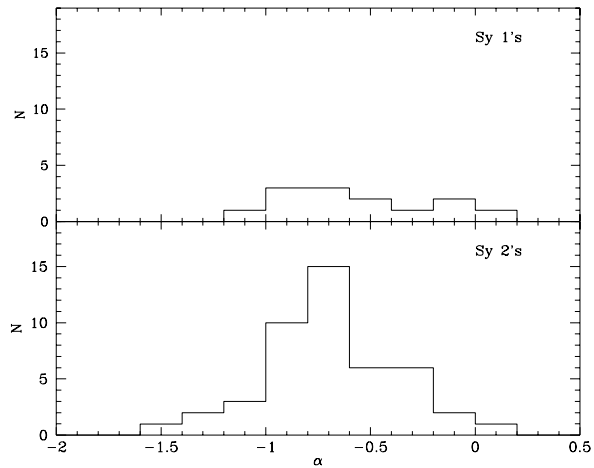,angle=0,width=9cm}}
\caption{ Histograms of the spectral index  distribution for Seyfert 1's and
Seyfert 2's.}
\end{figure}

\begin{figure}
\centerline{\psfig{figure=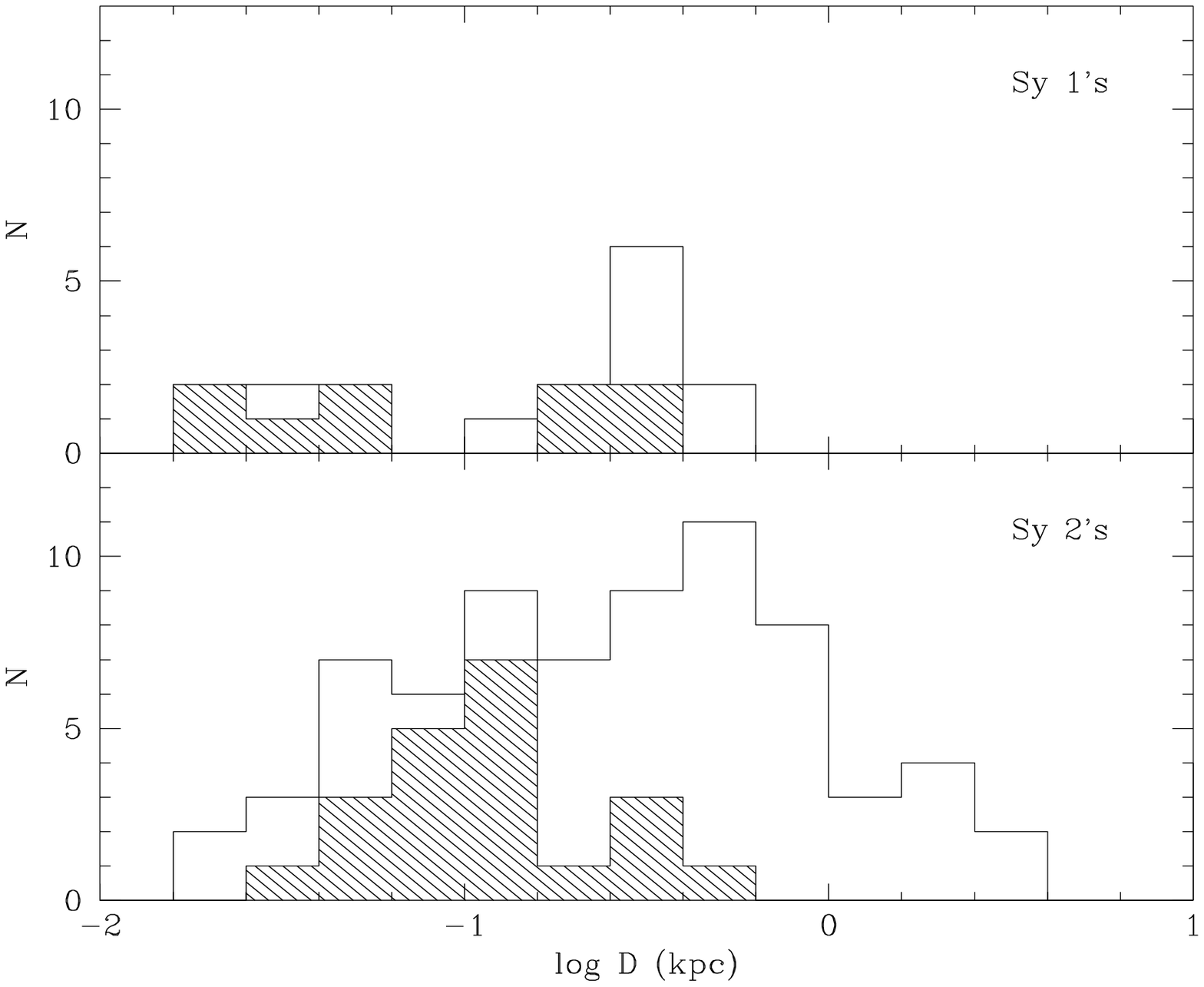,angle=0,width=9cm}}
\caption{ Histograms of the radio size distribution  for Seyfert 1's and
Seyfert 2's. Shaded regions represent upper limits.}
\end{figure}

\begin{figure}
\centerline{\psfig{figure=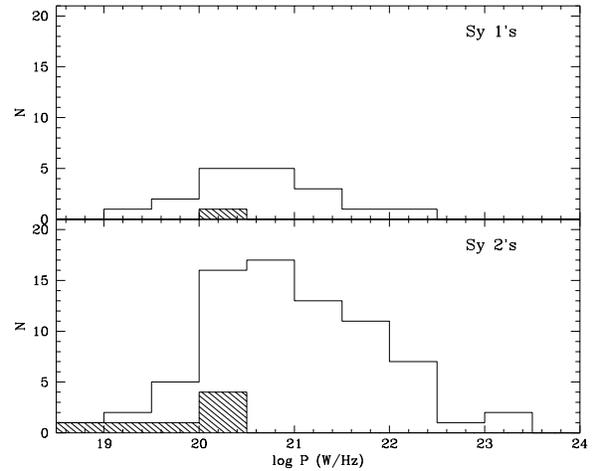,angle=0,width=9cm}}
\caption{ Histograms of the radio power distribution for Seyfert 1's and
Seyfert 2's. Shaded regions represent upper limits.}
\end{figure}

\subsection{A Comparison of Seyfert 1s and 2s}

Previous radio studies of volume-limited samples of Seyfert galaxies find: a)
marginal or no statistically significant difference between the distributions
of radio luminosities of Seyfert 1s and 2s; b) Seyfert~2's galaxies tend to
have a larger radio sources than Seyfert~1s, although at $<90$\% significance
level; c) the two types of Seyferts have essentially the same distribution of
spectral indices (\cite{giu90} 1990, Rush et al.  1996) although there seems
to be some evidence that flat spectrum cores are more common in Seyfert~1s
than in Seyfert~2s (\cite{ulv89}); and d) compact radio cores (on the
sub-arcsec scale) are more common in Seyfert~2s than in Seyfert~1s (R94). 

Examining first the distribution of spectral index (see histograms in
Fig.~17), we find a median value for the spectral index ($\alpha_6^{20}$ or
$\alpha_3^6$) of $\alpha = -0.44$ for Seyfert~1s and $-0.72$ for Seyfert~2s. 
However, the two distributions are not significantly different under a
Kolmogorov-Smirnov test (13\% probability that the two distributions are not
different).  Unfortunately, we have too few Seyfert 1's with measured spectral
indices to compare the frequency of flat-spectrum cores. 

We also compared the distribution of radio sizes and powers (see histograms in
Figs.~18 and 19 respectively) of Seyfert 1 and Seyfert 2 galaxies.  To improve
the fairness of the comparison, we only included sources out to a redshift of
$cz = 4000$~\kms , at which the redshift distributions of Seyferts 1 and 2
matched according to a K-S test.  At marginal significance, we find that
Seyfert 2 radio sources tend to be more luminous than Seyfert 1 radio sources;
the probability that Seyferts 1 and 2 arise from the same parent distribution
is $\sim 11\%$ for most two-sample tests, $\sim 25\%$ for the logrank test. 
Ulvestad \& Wilson (1984a) found a similar result for Markarian Seyferts, but
they showed in a follow-up paper (\cite{ulv89}) that the significance is
reduced owing to the paucity of low-luminosity Seyfert~2s in the Markarian
sample.  Inspection of the power-size diagram (Fig.~12) suggests that any
significance in the difference of radio powers owes to a handful of luminous
Seyfert~2s, but most Seyferts 1 and 2 have comparable radio powers.  Formally,
we measure mean log radio powers of $\log{P\ {\rm (W\ Hz^{-1})}} = 21.03\pm
0.01$ for Seyfert~2s, and $\log{P\ {\rm (W\ Hz^{-1})}} = 20.66\pm 0.18$ for
Seyfert~1s. 

In contrast, we measure a significant difference in radio sizes. All
of the two-sample tests report a difference in parent populations to a
significance of $\ge 95\%$, with the exception of the logrank test,
which is significant only to 87\%.  Seyfert 2s tend to be larger, with
a mean size of $0.53\pm 0.12$~kpc, compared to Seyfert 1s, mean size
$0.16\pm 0.04$~kpc. The errors on the mean sizes represent the dispersion
in the intrinsic value. Given the statistical agreement in the
distribution of radio powers, it is natural to interpret the size
difference in terms of an orientation unifying scheme. The prediction
of this model is the nuclear axes of Seyfert~1s are viewed more nearly
pole-on, and so radio jets are foreshortened by projection. 

\section{Conclusions}

We have presented VLA 6~cm and ATCA 3.5~cm radio data for 29 southern
Seyfert galaxies selected from a volume limited sample ($cz<3600$ km
s$^{-1}$).  The resolution of the observations is about $\sim
1$\arcsec, similar to that of the ground-based, optical narrow band
images.  The new radio observations presented in this paper cover more
than 50\% of our southern ($\delta<0\arcdeg$) volume limited
($cz<3600$ \kms) sample of Seyfert galaxies.  When data available from
the literature are added the coverage improves to greater than 80\%.
Only 2 of the 22 newly observed sources are undetected in our radio
observations.  Including data from the literature, we find 38\% of the
distance-limited sample is unresolved at arcsecond resolution, and
most of the resolved objects (34\% of the sample) show, as expected, a
linear structure.

Statistical comparisons are consistent with the picture that Seyferts
1 and 2 arise from the same parent population of AGN, but the radio
jets of Seyfert 1 galaxies are smaller in projection owing to their
orientation. This model is not contrary to the power-size correlation,
but, conversely, projection effects add noise to the measured
correlation. An additional prediction is that, binned in radio power,
Seyfert 1s should be smaller than Seyfert 2s. Unfortunately, owing to
the small number of sources per decade in radio power and the large
number of limits at low radio power, we are unable to test this
prediction with the current sample.

\acknowledgements

MA acknowledges the support of an Australian Post-Graduate Research
Award (APRA).  This research has made use of the NASA/IPAC
Extragalactic Database (NED) which is operated by the Jet Propulsion
Laboratory, Caltech, under contract with NASA.

\end{document}